\documentclass[fleqn,10pt,twoside]{article}

\usepackage[english] {babel}

\usepackage {graphicx}
\usepackage {graphics}
\usepackage {floatflt}
\usepackage {latexsym}
\usepackage {caption3} 
\usepackage[labelsep=period]{caption}[2007/11/04 v.3.1e]
\usepackage {amssymb}

\def\noi{\noindent}

\renewcommand{\thesubsubsection}%
        {\arabic{section}.\arabic{subsection}.\arabic{subsubsection}.}

\newcommand{\heads}[2]{\markboth{\protect\small\it #1}{\protect\small\it #2}}

\newcommand{\Arthead}[5]{ \setcounter{page}{#4}\thispagestyle{empty}\noi
    \unitlength=1pt \begin{picture}(500,40)

        \put(0,58){\shortstack[l]{\small\it Gravitation \& Cosmology,
                        \small\rm Vol. #1 (#2), No. #3, pp. #4--#5    \\
        \footnotesize {Proceedings of the 12th Russian Gravitational Conference, Kazan, 20-36 June 2005}    \\
\footnotesize\copyright \ #2 \ Russian Gravitational Society} }

    \end{picture}
	 }     		
\def\prepno#1#2
    {\thispagestyle{empty}
    \noi    \unitlength=1mm
    	\begin{picture}(178,10)
            \put(177,15){\llap{\bf #1}}
            \put(177,10){\llap{\small\rm #2}}
        \end{picture}
          }             

\newcommand{\Title}[1]{\noi {\uppercase{\Large #1}}     }
\newcommand{\Author}[2]{\noi{\large\bf #1}\\[2ex]\noindent{\it #2}   }

\newcommand{\Abstract}[1]{\vskip 2mm \begin{center}
        \parbox{16.4cm}{\small\noi #1} \end{center}\medskip}

\newcommand{\foom}[1]{\protect\footnotemark[#1]}

\newcommand{\email}[2]{\footnotetext[#1]{e-mail: #2}
		\addtocounter{footnote}{1}}

\heads{A.M.Baranov}
      {On an extension of a static ball solution}

\begin{document}
\twocolumn 
[
\Arthead{11}{2005}{4 ({\bf{44}})}{313}{314}

\Title{ON AN EXTENSION OF A STATIC BALL SOLUTION }


\vspace{.5cm}
   \Author{A.M.Baranov\foom 1 } 
{\it Dep. of Theoretical Physics,Krasnoyarsk State University,
79 Svobodny Prosp., Krasnoyarsk, 660041, Russia}

{\it Received 11 October 2005}

\Abstract
    {A new exact static interior solution of the Einstein equations is obtained for a gravitating ball filled with a Pascal perfect fluid . The solution is an extension of the well-known interior solution with a parabolic distribution of mass density and describes more compact astrophysical objects such as neutron and hyperon stars. It is shown that a behaviour of the new mass density distribution near the stellar centre can be described as a cusp catastrophe. }
]

\email 1 {alex\_m\_bar@mail.ru}

\section{Introduction}

The problem of finding exact solutions of Einstein's equations is connected, above all, with their nonlinearity.  Therefore every new exact solution always calls certain interest, both from the point of view of a physical interpretation and from the point of view of the method of its derivation. 

Schwarzschild's known interior solution (see [1]) is an exact interior solution of the Einstein's equations with a homogeneous distribution of a Pascal neutral perfect fluid. This solution describes the interior field of a static star. The main shortcoming of this solution is that the velocity of sound exceeds the velocity of light. The mass density is constant in the whole volume of the astrophysical object and has a finite jump on the stellar surface. 

Another exact interior solution [2],[3] is free of the above shortcoming. The solution describes a fluid ball with the parabolic law of mass density decrease from the centre to the surface. However, both solutions near the stellar centre behave almost equally with rrespect to Petrov's algebraic type. The Schwarzschild interior solution has the algebraic type 0. The second solution, in a small neighbourhood of the ball centre, has approximately the same algebraic type 0.

In the present article, the problem of generalization of the solution with a parabolic distribution of the mass density for a neutral Pascal fluid is considered.

\section{Metric and gravitational equations}

The Einstein equations with the static metric
$$ 
ds^2 = G^2(x)d\tau^2+2L(x)d\tau dx-x^2(d{\theta}^2+
sin{\theta}^2d{\varphi}^2) 
\eqno{(1)}
$$

\noindent
and the energy-momentum tensor of a perfect Pascal fluid 
$$
T_{\alpha\,\beta} = (\mu +p)\,u_{\alpha}\,u_{\beta} -p\,g_{\alpha\,\beta} 
\eqno{(2)}
$$
\noindent 
are considered. Thr function $\mu$ is the mass density and $p$ is the pressure; $u^{\alpha}=dx^{\alpha}/ds $ is the fluid 4D velocity; $\,\alpha\,,\,\beta\, = 0,1,2,3.$

These equations can be reduced to the nonlinear equation
of spatial oscillations
$$
G^{\prime \prime}_{\zeta \zeta} + {\Omega^2(\zeta(x))} G = 0,
\eqno{(3)}
$$

\noindent
where $ \zeta$ is the new variable  

$$
d\zeta =  dy/\sqrt {\varepsilon},
\eqno{(4)}
$$

\noindent
with $ \varepsilon = G ^ 2/L ^ 2 ;$  $\tau = t/R$ and $x = r/R $ 
are dimensionless variables; $0 \leq x \leq 1;$ $\;t, r$ are the time and radial coordinates, $ y=x^2 $, $0 \leq y \leq 1,$ and $ R $ is the stellar radius. The vacuum velocity of light is chosen to be equal to unity. 

We can write $ \Omega^2$ in a more convenient form: 

$$ 
\Omega^2 = - d(\Phi/y)dy
\eqno{(5)}
$$ 

\noindent
where 

$$
\Phi = 1-
\varepsilon = \displaystyle\frac{\chi}{2\sqrt{y}} \int \mu(y)\sqrt{y}dy = 
\displaystyle\frac{\chi}{x} \int \mu(x)x^2dx
\eqno{(6)}
$$

\noindent
plays the role of the Newton gravitational potential inside the ball; $\,\chi = \varkappa\,R^2.$

 When $\Omega^2$ equals zero, we have the Schwarzschild interior solution. For $\Omega^2 \equiv \Omega_0^2 = const,$ Eq.(3) is 
$$
G^{\prime \prime}_{\zeta \zeta} + \Omega_0^2 \, G = 0,
\eqno{(7)}
$$

\noindent 
and we have the known exact static solution [2], [3]
$$
G = \sqrt{g_{00}} = G_0\,cos(\Omega_0\,\zeta(x)+\alpha)
\eqno{(8)}
$$ 

\noindent
with $\Omega_0 = \chi\,\mu_0 a/5$ and the parabolic distribution of the mass density 

$$
\mu = \mu_0\,(1- a\,x^2)= \mu_0\,(1-a\,y),
\eqno{(9)}
$$ 

\noindent
where $a=(\mu_0-\mu_R)/\mu_0 \leq 1;\; \mu_0=\mu(x=0); \;\mu_R=\mu(x=1).$
 
This solution describes compact astrophysical objects such as neutron stars.

\section{Darboux's method and an extension of the solution }

According to Darboux's method [4], an extension of this solution is connected with the choice
$$
\Omega^2(\zeta)=\Omega_0^2 + w(\zeta),
\eqno{(10)}
$$
with $\Omega_0^2=const$ and 
$$
w(\zeta) = 2 (ln \tilde{Y})^{\prime \prime}_{\zeta \zeta},
\eqno{(11)}
$$ 
\noindent
where $\tilde{Y}$ is a particular solution of the equation of a linear
spatial oscillator (7) with $\tilde{\Omega}_0^2 = b^2 = const.$ 
Here we select the function $\tilde{Y} = cos(b\,\zeta)$. 

So we have the extension of Eq.(7)

$$
G^{\prime \prime}_{\zeta \zeta} + [\Omega_0^2 - 2b^2sec(b\,\zeta)]\,G = 0.
\eqno{(12)}
$$
The exact general solution of this Darboux equation is [5]
$$
  G(\zeta)=A \cdot [{\Omega_0} \cdot cos({\Omega_0}\,\zeta+\alpha_0) +
$$
$$
  + b \cdot tg(b\,\zeta) \cdot sin({\Omega_0}\,\zeta + \alpha_0)].
\eqno{(13)}
$$
The new function $\,\tilde{\Phi}(y)\,$ is written as a quadrature:
$$
\tilde{\Phi}(y) = (\chi\,\mu_0/3)y - \Omega_0^2\,y^2 + 
b^2 y \int{sec^2(b\,\chi(y))}dy.
\eqno{(14)}
$$
With the aid of this function we easily find the new mass density 
distribution 
$$
\chi \tilde{\mu}(y) = \chi \mu_0 -5\Omega_0^2\,y+6b^2 \int{sec^2(b\,\zeta(y)) dy}+
$$
$$
4b^2 y sec^2(b\,\zeta(y)),
\eqno{(15)}
$$
where $\tilde{\mu}(0)=\mu_0 .$ 

Now the new variable $\zeta$ is the quadrature
$$
\zeta(y) = \displaystyle{\frac{1}{2}\int{\frac{dy}{\sqrt{1-\tilde{\Phi}(y)}}}} = 
$$
$$
\displaystyle{\frac{1}{2}}\int{dy \Biggl[1-(\chi \mu_0/3)y +\Omega_0^2 y^2 }
$$

$$
-2b^2 y \int{sec^2 \Bigl( \displaystyle\frac{b}{2}\int{\displaystyle\frac{dy}{\sqrt{1-...}} } \Bigr )} \Biggr ]^{-1/2}   
\eqno{(16)}
$$

On the surface of the ball we have 
$$ 
\tilde{\Phi}(y=1) = \eta_0 + 2b^2 \equiv \eta^{*}, 
\eqno{(17)}
$$
where $\eta_0 = 2m/R$ is the compactness for the model with a parabolic distribution of mass density; $\,m\,$ is the integral mass of ball. 

The expression for the new mass density near the stellar centre  can be rewritten as 
$$
\displaystyle\frac{\tilde{\mu}}{\mu_0} = 1 - \displaystyle\frac{125 \sigma_0^2 - 784 b^2}{\sigma_0} x^2 + \displaystyle\frac{3 b^2}{2 \sigma_0} x^6 = 
$$
$$
= 1 - \displaystyle\frac{125 \sigma_0^2 - 784 b^2}{\sigma_0} y + \displaystyle\frac{3 b^2}{2 \sigma_0} y^3,
\eqno{(18)}
$$

\noindent
where $\sigma_0 = \chi \cdot \mu_0.$ 

This equation can be reduced to the equation which describes 
the cusp catastrophe 
$$
y^3 + (f_C - f)y +B\varepsilon_1 = 0,
$$
 \noindent
where $\varepsilon_1 = 1-\tilde{\mu}(y)/\mu_0 \ll 0$ is a small deviation from the mass density at the stellar centre, $f_C = 20/3,$ $\; B = (3/2)b^2/f\sigma_0$. 
 
The behaviour of the function $\tilde{\mu}(x)$ describes the star's stability near of the centre under the control parameters $\mu_0, b, \chi.$ There exists a range of the control parameters in which the system has a catastrophic behaviour.

\section{Summary}

Application of Darboux's  method to a known exact static interior solution of the gravitational equations with a parabolic mass density distribution makes it possible to obtain a new interior solution with a new mass density distribution law.  The extension of solution can be used for simulation of more compact astrophysical objects: neutron stars and, probably,  hyperon stars. Another feature of the new solution is that near the stellar centre the mass density behaviour is described by a cusp catastrophe, and, for some values of the parameters included in the expression for the mass density, the stability of the star breaks down.

\end{document}